\documentclass[twocolumn,twoside,slac_two]{revtex4}
\usepackage{graphicx}
\usepackage{fancyhdr}
\usepackage{longtable}
\begin{document}

%Title of paper
\title{The VERITAS Dark Matter Program}

% Repeat the \author .. \affiliation  etc. as needed
%
% \affiliation command applies to all authors since the last
% \affiliation command. The \affiliation command should follow the
% other information

\author{B. Zitzer}
\affiliation{Argonne National Laboratory, Argonne, IL, 60439 USA}
\author{for the VERITAS Collaboration}
\affiliation{veritas.sao.arizona.edu}

\begin{abstract}

In the cosmological paradigm, cold dark matter (DM) 
dominates the mass content of the Universe and is present 
at every scale. Candidates for DM include many extensions 
of the standard model, with a weakly interacting massive 
particle (WIMP) in the mass range from $\sim$10 GeV to 
greater than 10 TeV. The self-annihilation or decay of 
WIMPs in astrophysical regions of high DM density can 
produce secondary particles including very high energy 
(VHE) gamma rays with energy up to the DM particle mass. 
VERITAS, an array of atmospheric Cherenkov telescopes, 
sensitive to VHE gamma rays in the 85 GeV - 30 TeV energy 
range, has been utilized for DM searches. 
The possible astrophysical objects considered to be 
candidates for indirect DM detection are VERITAS dwarf 
spheroidal galaxies (dSphs) of the Local Group and the 
Galactic Center, among others. This presentation reports on
our extensive observations of these targets and constraints 
of the dark matter physics from these objects, including 
the methodology and preliminary results of a combined DM 
search of five dSphs.

\end{abstract}

%\maketitle must follow title, authors, abstract
\maketitle

\thispagestyle{fancy}

% body of paper here - Use proper section commands
% References should be done using the \cite, \ref, and \label commands
% Put \label in argument of \section for cross-referencing
%\section{\label{}}

\section{INTRODUCTION}

The search for Standard Model (SM) particles resulting from the
annihilation of Dark Matter particles provides an important complement to that of direct searches for DM interactions
and accelerator production experiments. Among the theoretical
candidates for the DM particle \cite{2005PhR...405..279B}, weakly interacting
massive particles are well motivated since they naturally
provide the measured present day cold DM density
\cite{2011ApJS..192...18K}. Candidates for WIMP dark matter are present in many
extensions of the SM of particle physics, such as supersymmetry
(SUSY) \cite{1996PhR...267..195J} or theories with extra dimensions \cite{2003NuPhB.650..391S}. In
such models, the WIMPs either decay or self-annihilate into
standard model particles, most of which produce either a
continuum of gamma rays with energies up to the DM particle mass, or mono-energetic gamma-ray lines.

Attractive targets for indirect DM searches are nearby
massive objects with high inferred DM density which are
not expected to be sources of VHE gamma rays. The Galactic
Center is likely the brightest source of gamma rays resulting from
DM annihilations, however the detected VHE gamma-ray emission
is coincident with the supermassive black hole Sgr A*
and a nearby pulsar wind nebula \cite{2006PhRvL..97v1102A}, motivating searches
for DM annihilation in the Galactic Center halo where the
VHE gamma-ray background is expected to be significantly lower \cite{2011PhRvL.106p1301A}. 
Galaxy Clusters have a large DM content. However, they are 
extended for VERITAS, and the possibility exists of a VHE 
background from conventional processes \cite{2004A&A...413...17P} \cite{1987A&A...182...21S}, 
although not yet detected. Galactic DM sub-halos would 
appear as unidentified objects (UNIDs) without multi-wavelength counterparts in Fermi-LAT data. If a Fermi UNID 
were detected in VHE, it could potentially be from DM. 
Dwarf spheroidal galaxies (dSphs) are additional attractive
targets for DM searches. Dwarf spheroidal galaxies
are relatively close ($\sim$50 kpc), and have a low rate of
active or recent star formation, which suggests a low 
background from conventional astrophysical VHE processes 
\cite{1998ARA&A..36..435M}. 

The following sections describe the status of observations 
and data analysis of each of the DM targets described above 
as of fall 2014, followed by conclusions and a discussion 
of the future of the VERITAS DM program. 

\section{SUB-HALOS}

\begin{table*}[t]
\begin{center}
\caption{Preliminary DM Sub-halo Results. Flux upper limits are given in units of Crab Nebula flux.}
\begin{tabular}{|l|c|c|c|c|c|}
\hline \textbf{2FGL Name} & \textbf{Exposure (hrs)} & \textbf{Significance ($\sigma$)} & \textbf{Excess Counts} &\textbf{$E_{tr}$ (GeV)} & \textbf{$F^{99\%CL}_\gamma (C.U.)$} \\
% & \textbf{(hrs)} & \textbf{($\sigma$)} & \textbf{(GeV)} & \textbf{($cm^{-2}s^{-1}$)} \\
\hline J0312.8+2013 & 9.7 & -1.5 & -26 $\pm$ 17 & 220 & $<$ 0.9\%    \\
\hline J0746.0$-$0222 & 9.1 & -0.9 & -15 $\pm$ 16 & 320 & $<$ 1.1\%    \\ 
\hline
\end{tabular}
\label{l2ea4-t1}
\end{center}
\end{table*}

Recent cosmological N-body, high-resolution simulations
\cite{springel} indicate that DM halos are populated
with a wealth of substructures \cite{diemand}. Because of 
tidal disruption near the Galactic disk, most of the
sub-halos are thought to survive at high Galactic latitude. 
The lack of material in these regions
prevents the DM overdensities from attracting enough 
baryonic matter and trigger star formation. DM
clumps would therefore be invisible to most astronomical 
observations from radio to X-rays. DM structures residing in the the Milky Way halo can be nearby the Sun and therefore
have a bright gamma-ray annihilation signal \cite{pieri08}. 
These clumps would likely be only visible at
gamma-ray energies and therefore may not have shown up in 
astronomical catalogs yet. Since
gamma-ray emission from DM annihilation is expected to be 
constant, DM clumps could then
appear in all-sky monitoring programs \cite{kamionkowski} done at gamma-ray energies. These can be best provided
by the Fermi-LAT instrument. Very likely, the distinct 
spectral cut-off at the DM particle mass
is located at energies too high to be measurable by Fermi 
within a reasonable timescale (see, e.g.
the WIMP mass lower limits in \cite{nakamura}) and can only be 
detected by ground-based telescopes, such as
VERITAS. Furthermore, detection of this spectral cut-off at 
the same energy in multiple objects
would stand as a visible signature of DM.
The Second Fermi-LAT Catalog (2FGL) contains 1873 high 
energy gamma-ray sources detected
by the LAT instrument after the first 24 months of 
observations. For each source, positional and
spectral information are provided as well as identification 
or possible associations with cataloged sources at other 
wavelengths. Although Fermi-LAT has a good angular 
resolution, a firm identification based on 
positional coincidence alone is not always feasible. Thus, 
576 sources in the 2FGL lack any clear association. These 
are the so-called unassociated Fermi objects (UFOs), a 
population among which DM clumps might be represented \cite{buckley}.
In order to extract possible DM clump candidates out of 
the 2FGL UNIDs, we adapt the selection criteria from
\cite{nieto}, selecting sources by requesting them:

\begin{itemize}

\item to lie outside the Galactic Plane,

\item to be non-variable,

\item to exhibit a power law spectra, and

\item to not have possible counterparts.

\end{itemize}

The original list obtained from the 2FGL catalog
is then filtered to select only sources observable
with VERITAS with a maximum zenith angle at culmination of 
40$^{\circ}$, in order to pursue the lowest energy 
threshold. Additionally, an estimate of required 
observation time for a 5$\sigma$ detection, dubbed 
\emph{detectability}, is computed based on a 2FGL Catalog 
flux extrapolation to the VERITAS energy range.

The preliminary results of the VERITAS observations shown 
in Table 1 are in tension with the extrapolation of 
the gamma-ray spectra from the Fermi-LAT to very high 
energies. Additional data from these UNIDs and others by VERITAS and other Cherenkov telescopes may completely completely rule out a direct extrapolation of the Fermi-LAT which would give strong DM model constraints or potentially detect a DM signature, provided that they are truly without counterparts at other wavelengths.

\section{GALAXY CLUSTERS}

Clusters of galaxies are the largest viralized objects in 
the Universe, with typical sizes of a few Mpc and masses on 
the order of 10$^{14}$ to 10$^{15}$ M$_{\odot}$. Most of 
the mass ($\sim80\%$) is dark matter, as indicated by 
galaxy dynamics and gravitational lensing 
\cite{2008SSRv..134....7D}. Aside from DM annihilation, it is 
possible to have gamma-ray emission from cosmic-ray 
interactions, producing neutral pions \cite{2004A&A...413...17P}, 
or inverse Compton of ambient photons 
\cite{1987A&A...182...21S}.  

VERITAS has taken 18.6 hours of dedicated observations of 
the Coma cluster between March and May 2008. The Coma 
cluster is a close (z=0.023) and massive (M$\sim10^{15}
M_{\odot}$) cluster which has been thoroughly studied across 
all wavelengths. The standard analysis of the Coma 
cluster data using point-source cuts for the core of the 
cluster yielded 17 excess counts, with a significance of 
0.84$\sigma$, indicating a non-detection. Upper limits of 
0.83\% of the Crab Nebula flux were placed for the core of 
the Coma cluster with 95\% confidence, assuming a powerlaw 
spectral index of -2.3. With the absence of a signal from 
the Coma cluster, limits of the velocity-averaged cross-
section for DM annihilation were placed at 
$\mathrm{{\cal{O}}(10^{-21})}$, as shown in Figure 1 \cite{2012ApJ...757..123A}.  

An archival VERITAS galaxy cluster search is also in the 
works, looking for galaxy clusters that have happened to 
overlap in the same FOV as other targeted observations. ROSAT and SDSS galaxy cluster catalogs are being used to cross-check with other VERITAS observations, Most notably M87 in the Virgo cluster of galaxies \cite{VerM87} and NGC 1275 in the Perseus cluster \cite{VerVirgoNGC1275}.  

\begin{figure}
\includegraphics[width=\columnwidth,clip=true,trim=0 2mm 0 0] {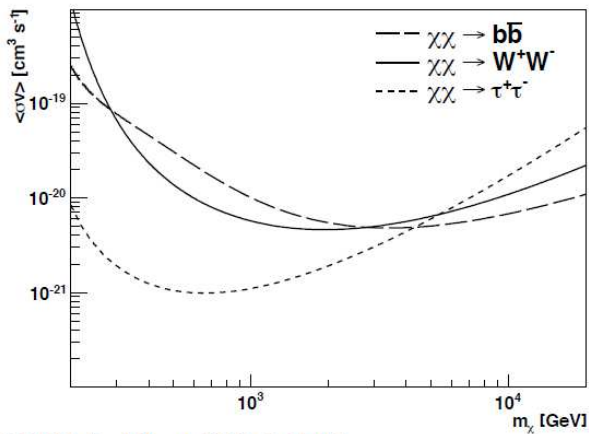}
\caption{Dark Matter velocity-averaged cross-section limits from the Coma galaxy cluster. Figure taken from \cite{2012ApJ...757..123A}.}
\end{figure}

\section{GALACTIC CENTER}

The center of our galaxy, SgrA*, is a strong VHE source, 
along with several other VHE sources nearby 
\cite{HESS_SgrA} and possible diffuse emission 
\cite{HESS_SgrA_Diffuse}, making Dark Matter detection in 
that region a complicated, but not impossible, prospect.  

The Galactic center was observed by VERITAS in 2010-2014 
for $\sim80 \, \rm{hrs}$ (good quality data) at zenith 
angles of $z = 60-66 \deg$ (average threshold of 
$E_{\rm{thr}} \simeq 2.5 \, \rm{TeV}$). The higher 
effective areas due to the large zenith angle observations 
make the VERITAS observations now the most sensitive 
instrument for the Galactic Center region above 2 TeV. The 
detection of SgrA* by VERITAS and VHE emission in the 
region through conventional processes are discussed in 
greater detail elsewhere in these proceedings 
\cite{SmithVerSgrA}. 

The DM signal and background regions for the Galactic center region will use 
arc-shaped regions north and south of the Galactic plane 
to avoid diffuse emission and VHE sources, as shown in 
Figure 2. The VERITAS observations were accompanied by 
off-source 
observations of a field located in the vicinity of the 
Galactic center region (with similar zenith angles
and sky brightness) without a known TeV $\gamma$-ray 
source. These observations are used to study the 
background acceptance throughout the field of view and 
will assist in the identification of a diffuse $\gamma$-ray
component surrounding the position of the Galactic center.
 
The DM search for the Galactic center region is still in the preliminary stages. Work is currently underway for computing J factors for the signal and background regions for the arc-shaped regions described above.  

\begin{figure}
\includegraphics[width=\columnwidth] {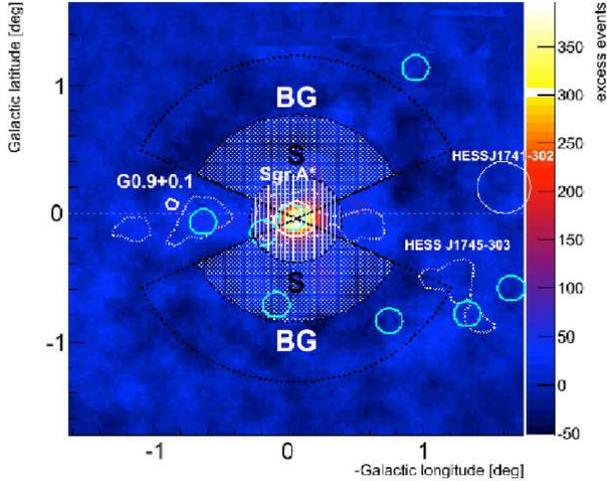}
\caption{Skymap of the galactic center region using a subset of the VERITAS data. DM signal and background regions are indicated north and south of SgrA*.}
\end{figure}

\section{DWARF GALAXIES}

\begin{table*}[t]
\begin{center}
\caption{Preliminary DM DSph Results}
\begin{tabular}{|l|c|c|c|c|c|}
\hline \textbf{DSph Name} & \textbf{Exposure (hrs)} & \textbf{Significance ($\sigma$)} & \textbf{Excess Counts} &\textbf{$E_{tr}$ (GeV)} & \textbf{$F^{99\%CL}_\gamma (C.U.)$} \\
% & \textbf{(hrs)} & \textbf{($\sigma$)} & \textbf{(GeV)} & \textbf{($cm^{-2}s^{-1}$)} \\
\hline Segue 1 & 92.0 & 0.7 & 94.4 $\pm$ 134.1 & 150 & $<$ 0.4\%    \\
\hline Ursa Minor & 59.7 & -0.1 & -7.2 $\pm$ 68.5 & 290 & $<$ 0.3\%    \\ 
\hline Draco & 49.9 & -1.0 & -73.2 $\pm$ 69.1 & 220 & $<$ 0.3\%    \\ 
\hline Bo{\"o}tes 1 & 14.0 & -1.0 & -38.5 $\pm$ 36.7 & 170 & $<$ 0.3\%    \\ 
\hline Willman 1 & 13.7 & -0.6 & -28.7 $\pm$ 46.2 & 180 & $<$ 1.0\%    \\ 
\hline
\end{tabular}
\label{l2ea4-t1}
\end{center}
\end{table*}

Dwarf spheroidal galaxies (dSphs) best meet the criteria 
for a clear and unambiguous detection of dark matter. They 
are gravitationally-bound objects and contain up to $
\mathrm{{\cal{O}}(10^{3})}$ times more mass in DM than in 
visible matter [1]. As opposed to the Galactic center, 
globular clusters and clusters of galaxies, dSphs present 
the clear advantage of being free of any significant 
astrophysical emission. Their high Galactic latitude and 
relative proximity to Earth ($\sim50$ kpc) make them very 
good targets for high signal-to-noise detection.

Between the start of full VERITAS array 
operations and Summer 2013, five dSphs have been 
observed with VERITAS: Segue 1, Ursa Minor (UMi), Draco, 
Bo{\"o}tes 1, and Willman 1. The VERITAS collaboration has previously 
published a 48 hour exposure on Segue 1 \cite{dSph2} 
and $\sim$15 hour exposures on the other four mentioned 
here \cite{dSph1}. Deeper exposures on Segue, UMi and 
Draco have been taken after these publications. To obtain 
the lowest possible energy threshold for DM searches, 
looser cuts optimized $\it{a pirori}$ on soft spectral VHE sources were 
used for the collective data set. The combination of looser cuts and deeper exposures 
revealed certain systematic effects in the cosmic-ray (CR) background data. The 
first is a gradient in the VERITAS cameras dependent on the 
zenith angle of observations across the FOV. The second 
systematic effect results from bright stars in the VERITAS FOV 
that cause the high voltage to pixels in the cameras to be suppressed. Both of these 
systematic effects have been corrected for and the results are 
summarized in Table 2. The `crescent' background method which also developed for dSph analysis \cite{2013arXiv1307.8367Z} was also used for the Table 2 results. 

The first background systematic effect, relating to the zenith 
gradient, was corrected using a zenith-dependent acceptance 
map. The standard VERITAS analysis uses only a radially-dependent 
acceptance, i.e. the angle between the reconstructed 
event direction and the array pointing direction. Measuring the 
gradient utilized a map that is the ratio of the number of 
all reconstructed events in a sky map within a given search radius (defined as 0.17 degree in this work) to the (radial only) 
acceptance in that same bin, a parameter we will refer to 
as $\it{flatness}$ in the rest of this work. If the 
acceptance adequately describes the CR background, then 
excluding any stars or known VHE sources, it should not 
correlate with any external parameters. However, a strong 
correlation was seen with the mean zenith angle of each 
reconstructed event position in the skymap bin. This 
correlation in the skymap bins is shown in Figure 3. This gradient is 
corrected in the data by fitting the correlation with a 
fourth-degree polynomial and using that to re-weight the 
acceptance map. The $\alpha$ parameter from Li \& Ma 
equation 17 is then re-calculated \cite{1983ApJ...272..317L}, 
\cite{2007A&A...466.1219B}. It should be noted that 
the difference of the adjusted value of $\alpha$ to the non-adjusted value is typically less than 1\%. However, the 
difference to the $\gamma$-ray excess and significance 
becomes larger over time as statistics accumulate.  

\begin{figure}
\includegraphics[width=\columnwidth] {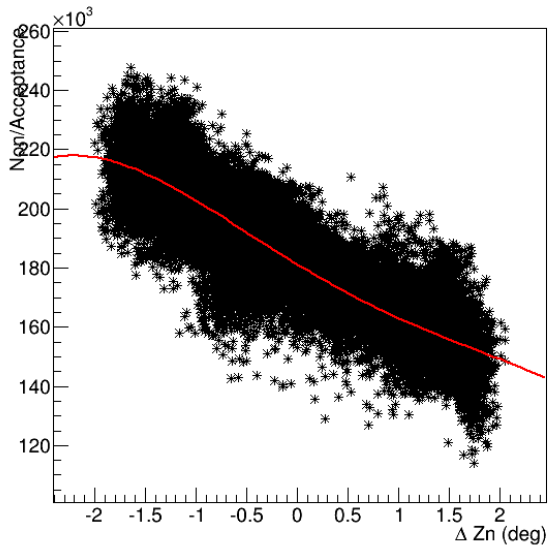}
\caption{Scatter plot of the flatness parameter on the y-axis, and the mean zenith difference between the array tracking direction and event reconstruction direction on the x-axis for the Segue 1 data summarized in Table 2. A fit of this scatter to a fourth-degree polynomial is shown in red.}
\end{figure}

\begin{figure}[h]
\includegraphics[width=\columnwidth] {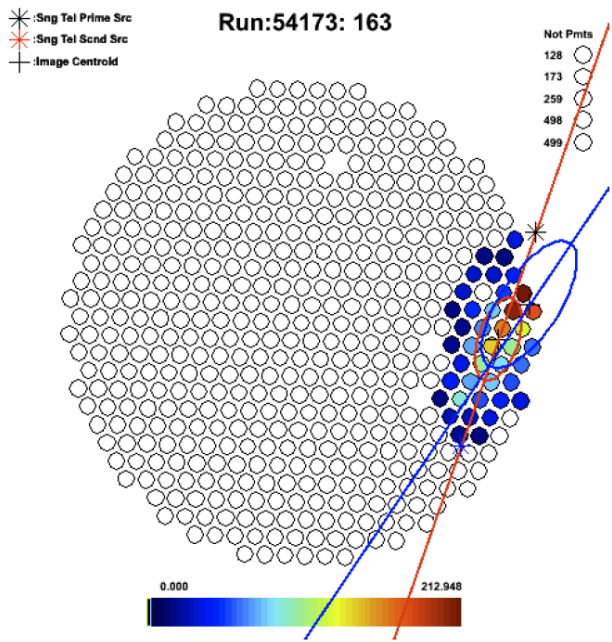}
\caption{Example of the HFit and standard Hillas event characterization. The HFit shower image is the blue ellipse outline, while the standard Hillas moment analysis is the red ellipse outline. Figure from \cite{2012AIPC.1505..709C}.}
\end{figure}

\begin{figure}[h]
\includegraphics[width=\columnwidth,clip=true,trim=0 0 0 5mm] {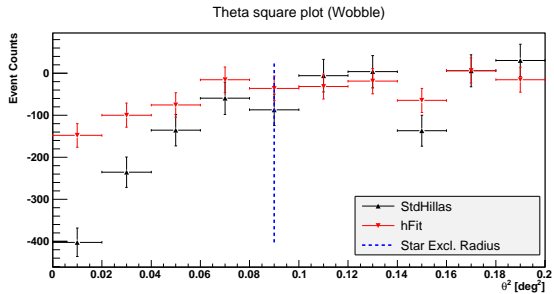}
\caption{Apparent surface brightness of background cosmic ray events as a function of angle $\theta$ from the bright star, Merak, in the FOV of RGB J1058+564 using both the HFit algorithm and the standard Hillas shower image characterization. The VERITAS data analysis typically excludes regions around bright stars or known VHE sources from cosmic-ray background characterization. The default radius for background exclusion region for Merak is also shown as the dashed blue line. }
\end{figure}

The second background systematic effect is due to ``holes" that are 
seen in the data relating to bright stars in the FOV that 
would trip the high voltage of camera pixels or raise cleaning thresholds 
due to higher night-sky background levels. Missing pixels would in turn effect both energy and gamma-ray position reconstruction. A method 
of using a 2D Gaussian likelihood fit to each shower image 
is utilized here, called \emph{HFit} \cite{2012AIPC.1505..709C}. Standard VERITAS 
analysis uses the moments of the shower images, commonly 
referred to as Hillas parameters. By 
using the 2D elliptical fitting to each image, missing data 
from disabled or broken PMTs are effectively interpolated 
around, as are images truncated by the edge of the cameras, 
as shown in Figure 4. This has been shown to reduce both 
the size and the depth of the holes due to bright stars 
seen in the data, including but not limited to the 
B magnitude 3.4 star Eta Leonis which is located 
0.68$^{\circ}$ away from the center of the Segue 1 dSph. 
The effectiveness of HFit on a independent data sample is 
shown in Figure 5, which shows the apparent surface brightness in the CR background (which 
is in reality a deficit for reasons described above) at a star location in the FOV of 
the blazar RGB J1058+564 (Merak, 2.4 B magnitude).

Work is currently underway to utilize the data for the 
previously published papers plus additional data for a 
combined DM physics result. This result will use the 
methodology developed by Geringer-Sameth et al. 
\cite{2014arXiv1410.2242G} to utilize both the 
individual energy and event reconstruction information as well as astrophysical 
``J factors" from a generalized NFW profile by Geringer-
Sameth and Walker \cite{2014arXiv1408.0002G}.

\section{CONCLUSIONS}

New DM publications from the VERITAS collaboration are 
forthcoming: the 
combined analysis of the dSphs should be publically 
available within the next six months, which promises to be 
the most robust result of any DM result in VHE gamma rays so far, while 
DM results for the Galactic Center, Fermi UNIDs and the 
archival galaxy cluster search should be ready on longer 
timescales. New analysis techniques are being developed by the VERITAS collaboration which promise large gains to our DM sensitivity, as an example an extended analysis of the dSphs which would incorporate longer tails of the DM density profile, which would in turn give a boost to the J factors by as much as a factor of $\sim$2. A combined analysis with Fermi-LAT, or other $\gamma$-ray instruments could potentially provide a boost to DM sensitivity. 

The VERITAS collaboration has a historical commitment to substantial DM observations and 
plans to do so in the foreseeable future. Recently, a new 
long-term plan for VERITAS has gone into effect, which has 
a significant fraction (15-20\%) of the total dark observation time dedicated to DM observations. The focus of this long-term plan is dSphs; however the galactic 
center, Fermi UNIDs and galaxy clusters will not be 
completely ignored. If executed consistently over the 
expected lifetime of VERITAS, these observations will form 
the basis of an important and unique contribution 
to the field of indirect DM detection. 

\bigskip % extra skip inserted
\begin{acknowledgments}

This research is supported by grants from the U.S. 
Department of Energy Office of Science, the U.S. National 
Science Foundation and the Smithsonian Institution, by 
NSERC in Canada, by Science Foundation Ireland (SFI 10/RFP/
AST2748) and by STFC in the U.K. We acknowledge the 
excellent work of the technical support staff at the Fred 
Lawrence Whipple Observatory and at the collaborating 
institutions in the construction and operation of the 
instrument. The VERITAS Collaboration is grateful to Trevor Weekes for his seminal contributions and leadership in the field of VHE gamma-ray astrophysics, which made this study possible. 

\end{acknowledgments}

\bigskip % extra skip inserted
% Create the reference section using BibTeX:
%\bibliography{basename of .bib file}
%\begin{thebibliography}{9}   % Use for  1-9  references

\end{document}